\documentclass[
showpacs,
floatfix,
aps,
prl,
twocolumn,
superscriptaddress,
amssymb,
]{revtex4-1}

\usepackage{amssymb}
\usepackage{latexsym}
\usepackage[dvips]{graphicx}
\usepackage{amsmath}

\usepackage{graphicx}
\usepackage{dcolumn}
\usepackage{amsfonts}
\usepackage{bm}
\usepackage{epsfig}

\newcommand{\be}{\begin{equation}}
\newcommand{\ee}{\end{equation}}
\newcommand{\bea}{\begin{eqnarray}}
\newcommand{\eea}{\end{eqnarray}}

\def\bip{B_\parallel}

\def\ar{\rho_{xx}/\rho_{yy}}
\def\rxx{R_{xx}}
\def\ryy{R_{yy}}
\def\rhoxx{\rho_{xx}}
\def\rhoyy{\rho_{yy}}

\newcommand{\rfig}[1]{Fig.\,\ref{#1}}
\newcommand{\rFig}[1]{Figure \,\ref{#1}}

\newcommand{\rref}[1]{Ref.\,\onlinecite{#1}}

\begin{document}
\title{Transport anisotropy in Ge quantum wells in the absence of quantum oscillations}
\author{Q.~Shi}
\affiliation{School of Physics and Astronomy, University of Minnesota, Minneapolis, Minnesota 55455, USA}
\author{M.~A.~Zudov}
\email[Corresponding author: ]{zudov@physics.umn.edu}
\affiliation{School of Physics and Astronomy, University of Minnesota, Minneapolis, Minnesota 55455, USA}
\author{C.~Morrison}
\affiliation{Department of Physics, University of Warwick, Coventry, CV4 7AL, United Kingdom}
\author{M.~Myronov}
\affiliation{Department of Physics, University of Warwick, Coventry, CV4 7AL, United Kingdom}

\begin{abstract}
Recent study of a high-mobility 2D hole gas in a strained Ge quantum well revealed strong transport anisotropy in the quantum Hall regime when the magnetic field was tilted away from the sample normal.\citep{shi:2015a} 
In the present study we demonstrate that the anisotropy persists to such high temperatures and filling factors that quantum oscillations are no longer observed. 
This finding rules out the formation of a stripe phase as a possible origin for the observed anisotropy.
However, we also show that the observed anisotropy is not consistent with other known anisotropies, such as those arising from finite thickness effects or surface roughness.
\end{abstract}
\maketitle


It is well established that transport properties of 2D systems could be modified by a pure in-plane magnetic field $B = \bip$ for several reasons.
First, $\bip$ can align the spin of the charge carriers leading to an increase of the resistivity due to suppression of screening by charged impurities \citep{tutuc:2002,dassarma:2005a}.
Second, due to a finite thickness of a 2D system, $\bip$ distorts the Fermi contour and modifies the scattering rates, also producing positive magnetoresistance \citep{heisz:1996,dassarma:2000,papadakis:2000,zhou:2010}.
Finally, the increase of the resistivity with $\bip$ could also occur because of interface roughness \citep{bykov:2001a,bykov:2001b,goran:2008}, due to local, anisotropic perpendicular magnetic fields.
Both finite thickness and roughness mechanisms imply some anisotropy in the resistivity tensor, albeit with different orientations of the anisotropy axis with respect to $\bip$.
The spin-polarization scenario, on the other hand, does not lead to anisotropy, unless the crystal structure is anisotropic.
An addition of a weak perpendicular magnetic field ($B_\perp \ll \bip$) can further modify the in-plane magnetoresistance and the anisotropy.
However, if one disregards the appearance of quantum oscillations, the effect of $B_\perp$ is usually rather small \citep{heisz:1996,englert:1983,streda:1995,kamburov:2012,kuntsevich:2013,kamburov:2013b}.

In a purely perpendicular magnetic field, $B = B_\perp$, 2D systems reveal a much wider variety of transport phenomena.
At low $B$, these phenomena include several kinds of both positive and negative\citep{paalanen:1983,choi:1986,li:2003,bockhorn:2011,hatke:2012a,shi:2014a,shi:2014c,bockhorn:2014} magnetoresistances, which can originate from electron-electron interactions\citep{altshuler:1985,girvin:1982,gornyi:2003,gornyi:2004} or quasiclassical memory effects \citep{baskin:1978,bobylev:1995,fogler:1997,dmitriev:2001,dmitriev:2002,mirlin:2001,polyakov:2001}.
At higher $B$, much more dramatic phenomena, such as integer\citep{klitzing:1980} and fractional\citep{tsui:1982} quantum Hall (QH) effects, stripe and bubble phases,\citep{koulakov:1996,lilly:1999a,du:1999,fradkin:1999} as well as Wigner crystals,\citep{wigner:1934,lozovik:1975,lam:1984,levesque:1984,chen:2006} emerge due to interplay among Landau/Zeeman quantizations, disorder, and electron-electron interactions.
Added $\bip$ can significantly change the transport properties owing to, e.g., spin polarization,\citep{nicholas:1988,du:1993,leadley:1998} modification of scattering rates \citep{gusev:1999,hatke:2011a,bogan:2012}, and finite thickness effects \citep{hatke:2012c}.
Unless already anisotropic, the system remains isotropic with few exceptions, such as a $\bip$-induced stripe phase in the $N = 1$ Landau level \citep{lilly:1999b,pan:1999}.

It was recently realized that when a high-mobility 2D hole gas (2DHG) in a strained Ge quantum well is subject to \emph{both} the in-plane ($B_x = B \sin  \theta$) and the \emph{out-of-plane} ($B_z =  B \cos \theta$) magnetic fields, its low temperature transport properties in the QH regime become strongly anisotropic.\citep{shi:2015a}
At $T \approx 0.3$ K, and $B_z$ larger than the onset of spin-splitting, the resistivity ratio at half-integer filling factors was found to increase gradually with $\theta$, reaching $\ar\approx 11.5$ at $\theta = 80^\circ$.
At smaller $B_z$, the anisotropy decreased roughly linearly with $B_z$ for \emph{all} $\theta$, until vanishing close to the onset of Shubnikov-de Haas oscillations.
Finally, switching off \emph{either} $B_z$ \emph{or} $B_x$ resulted in a roughly isotropic state with $\ar \approx 1$ over a wide range of $B_x$ or $B_z$ (up to 7 T).

The observed anisotropy was examined in terms of a stripe/nematic phase,\citep{koulakov:1996,lilly:1999a,du:1999,fradkin:1999} known to occur in high ($2 \le N \le 6$) Landau levels of ultra-clean GaAs systems cooled down to $T\lesssim 0.1$ K \citep{lilly:1999a,du:1999}.
While some features were consistent with the stripe scenario, slow decay of the anisotropy with $T$ seemed to rule against it.
As the focus of \rref{shi:2015a} was on near half-integer filling factors for $4 < \nu < 40$ in the QH regime, measurements were limited to $T < 1.5$ K and moderate $B_x$, which implied $B_x/B_z < 6$.
It is thus important to investigate if the anisotropy can survive at higher $T$ and higher tilt angles when the quantum oscillations are absent. 
It is also interesting to extend the study to the lower $N<2$ Landau levels, where the nematic phases in GaAs are less likely to occur.

In this article we report on transport measurements in a high-mobility 2DHG in a Ge quantum well in tilted magnetic fields up to 18 T, focusing on the regime of (i) much higher $B_x/B_z$ and $T$ up to 8 K and (ii) the $N$ = 1 Landau level.
We find that while the anisotropy smoothly increases with $B_x$, addition of a small perpendicular magnetic field $B_z \lesssim 0.5$ T significantly enhances the anisotropy without bringing in quantum oscillations.
At $B_z \gtrsim 0.5$ T, we find that the main result of \rref{shi:2015a}, namely that $\ar$ is determined by the tilt nagle alone, holds all the way up to $B_x/B_z \gtrsim 20$ and to much higher $T$, even in the absence of quantum oscillations.
The existence of the anisotropy in the regime where no quantum oscillations are seen allows us to rule out the formation of stripe phase as a possible origin.
We further demonstrate that our findings are not compatible with other known anisotropies, such as those arising from finite-thickness effects\citep{dassarma:2005a} or surface roughness \citep{bykov:2001a,bykov:2001b,goran:2008}, pointing towards a novel mechanism of anisotropic transport.
We also find that at low temperatures and at fixed tilt angle, the anisotropy is significantly suppressed in the $N$ = 1 Landau level, indicating that the ``scaling'' of the anisotropy with the tilt angle breaks down.

Our sample is a $5\times5$ mm square fabricated from a fully strained, $17$ nm-wide Ge quantum well grown by reduced pressure chemical vapour deposition on a relaxed Si$_{0.16}$Ge$_{0.84}$/Ge/Si(001) virtual substrate \citep{dobbie:2012,zudov:2014,morrison:2014,myronov:2014,shi:2014b,myronov:2015,shi:2015c}.
Holes are supplied by a 12 nm-wide B-doped layer separated from the interface by a 30 nm-wide undoped Si$_{0.16}$Ge$_{0.84}$ spacer. 
At $T$ = 0.3 K, our 2DHG has density $p \approx 2.9 \times 10^{11}$ cm$^{-2}$ and mobility $\mu \approx 1.3 \times 10^6$ cm$^2$/Vs. 
The resistances $\rxx$ and $\ryy$ were measured using corner contacts by a low-frequency (a few Hz) lock-in technique.
The sample was mounted on a rotator stage and the angle between the sample normal and the magnetic field was could be changed \emph{in situ} without warming up the sample.
Magnetotransport measurements were done by either sweeping magnetic field at a fixed angle or rotating the sample in a fixed magnetic field.

\begin{figure}[t]
\includegraphics{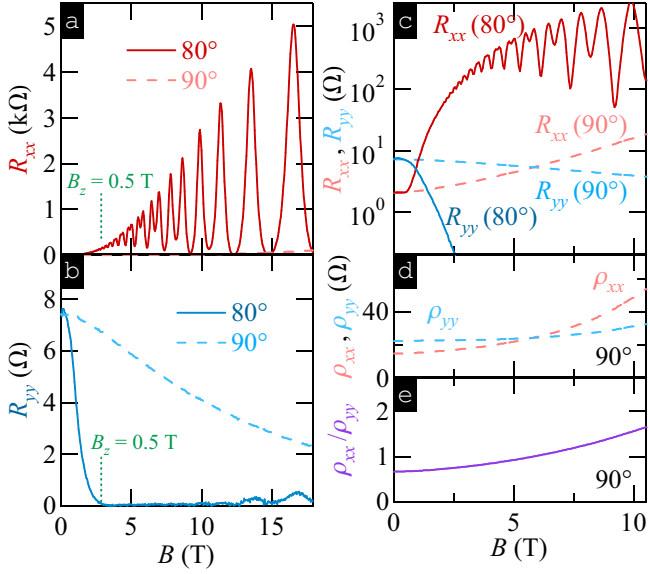}
\caption{(Color online)
(a) $\rxx(B)$ [in k$\Omega$] and (b) $\ryy(B)$ [in $\Omega$] at $\theta = 80^{\circ}$ (solid lines) and $90^{\circ}$ (dashed lines) at $T \approx 0.3$ K.
(c) same as above on a log-linear scale.
(d) $\rhoxx(B)$, $\rhoyy(B)$ and (e) $\ar(B)$ at $\theta = 90^\circ$.
}
\label{fig1}
\end{figure}

In \rfig{fig1} we compare magnetoresistances (a) $\rxx(B)$ and (b) $\ryy(B)$ measured in a parallel field ($\theta = 90^{\circ}$, $B=\bip=B_x$, dashed line) to their values in tilted field ($\theta = 80^{\circ}$, $B  \approx 1.015\,B_x$, solid line).
All four traces shown in \rfig{fig1}(a) and \rfig{fig1}(b) are also presented in \rfig{fig1}(c) on a log-linear scale.
At $B = 0$, our 2DHG exhibits modest anisotropy with $\rxx < \ryy$, which likely originates from anisotropic surface roughness.\citep{hassan:2014}
This anisotropy virtually disappears upon application of a purely perpendicular magnetic field $B = B_z > 0.1$ T \citep{shi:2015a}.
If a purely parallel field is applied, $B = B_x$ ($\theta = 90^\circ$), $\rxx$ increases, $\ryy$ decreases, and at $B_x \approx 6$ T one finds $\rxx \approx \ryy$.
On the other hand, when a small perpendicular field is added ($\theta = 80^\circ$), both $\rxx$ and $\ryy$ show much bigger changes starting from $B_x \approx 0.5$ T and differ by three orders of magnitude at $B_x \approx 2.8$ T.
This value of $B_x$ corresponds to $B_z = 0.5$ T, marked by dotted vertical line. 

Since $\rxx$ and $\ryy$ are measured in a square sample, the decrease of $\ryy$ doesn't necessarily mean the decrease of resistivity $\rhoyy$.
Following the results of \rref{simon:1999} we convert $\rxx,\ryy$ to $\rhoxx,\rhoyy$ and present the results versus $B_x$ at $\theta = 90^\circ$ in \rfig{fig1}(d).
We find that $\rhoyy$ increases slower than $\rhoxx$, and the resistivity ratio becomes $\ar \approx 1.6$ at $B = B_x$ = 10 T, as illustrated in \rfig{fig1}(e).
We thus confirm that a purely parallel magnetic field produces only a modest transport anisotropy.

\begin{figure}[t]
\includegraphics{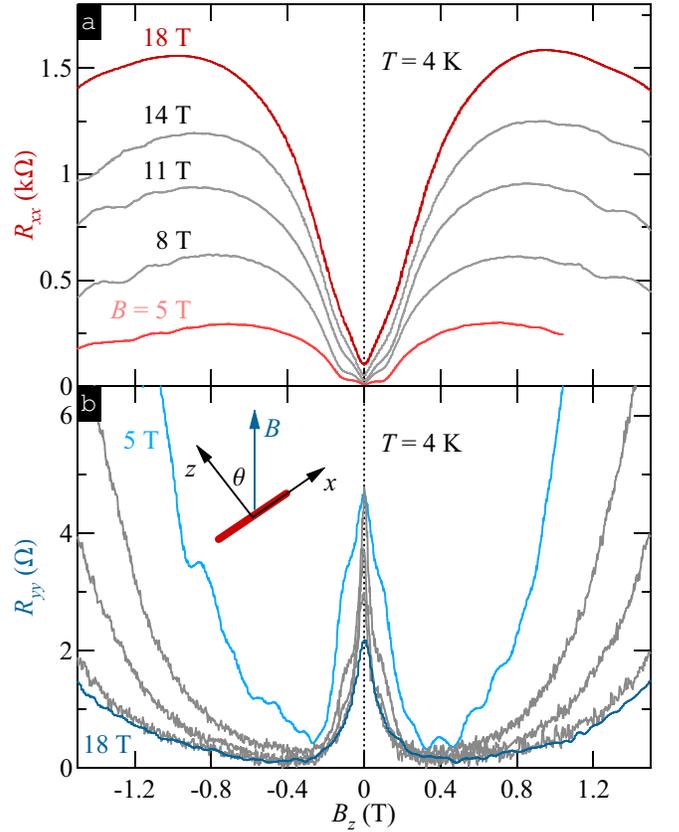}
\caption{(Color online)
(a) $\rxx$ [in k$\Omega$] and (b) $\ryy$ [in $\Omega$] measured at $T = 4$ K and different $B$, as marked, versus $B_z$, introduced via rotation of the sample.
}
\label{fig2}
\end{figure}

To examine the anisotropy in the regime when quantum oscillations are absent, we perform the transport measurements at elevated temperature of $T = 4$ K and at large tilt angles.
To access the high angle limit, we apply a fixed magnetic field $B$ along $\hat{x}$ direction and then rotate the sample about $\hat{y}$-axis to introduce a small perpendicular field $B_z$.
In \rfig{fig2} we present (a) $\rxx$ and (b) $\ryy$, measured at $T  = 4$ K, versus $B_z$, introduced via rotation of the sample in different $B \approx B_x$ from 5 to 18 T, as marked.
We observe that $\rxx$ ($\ryy$) initially increases (decreases) with $B_z$ and then shows a maximum (minimum) at all $B$ studied.
With increasing $B\approx B_x$, the maximum (minimum) becomes higher (lower) and gradually shifts to higher $B_z$.
Based on these observations we conclude that the anisotropy (a) can be significant even at $T = 4$ K, (b) does not require quantum oscillations, and (c) monotonically increases with $B_x$ while exhibiting a maximum at $B_z$ somewhere between $0.2$ and $0.8$ T.

\begin{figure}[t]
\includegraphics{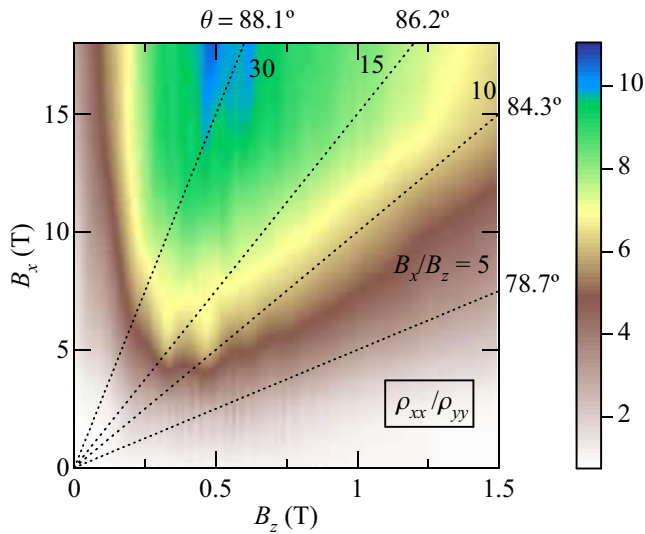}
\caption{(Color online)
$\ar$ versus ($B_z,B_x$) at $T = 4$ K.
Dotted lines are drawn at $B_x/B_z = 5, 10, 15$, and 30.
}
\label{fig3}
\end{figure}

Having determined the range of $B_z$ where the anisotropy is maximized, we present in \rfig{fig3} a false color plot of $\rhoxx/\rhoyy$ versus $B_z$ and $B_x$.
The strongest anisotropy, characterized by $\ar >$ 10, occurs at $B_x \gtrsim 10$ T in a region which is domed at $B_z \approx 0.5$ T. 
This dome has a considerably larger gradient on the lower $B_z$ side than at the higher $B_z$ side.
Furthermore, the iso-anisotropy lines on the higher $B_z$ side are well described by constant $B_x/B_z$, as illustrated by dotted lines.
While this result was already obtained for half-integer filling factors in the QH regime at $T \approx 0.3$ K and $B_x/B_x < 5.7$ \citep{shi:2015a}, here we demonstrate that the same rule applies for much higher $T$, in the regime where there are no quantum oscillations, and up to much higher $B_x/B_z$.
However, this rule breaks down on the other side of the dome, where, as we show next, the anisotropy is controlled by the perpendicular component of the magnetic field.

\begin{figure}[t]
\includegraphics{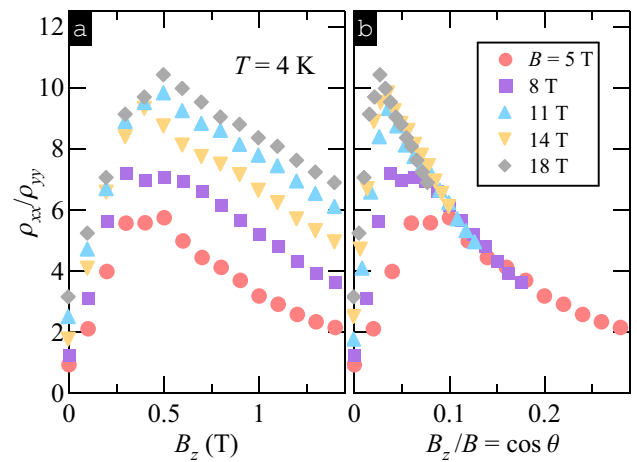}
\caption{(Color online)
$\ar$ versus (a) $B_z$ (b) $B_z/B$ at $B_x \approx B = 5, 8, 11, 14$, and 18 T at $T = 4$ K.
}
\label{fig4}
\end{figure}

In \rfig{fig4}(a) we present $\ar$  versus $B_z$ measured at different $B_x \approx B$, as marked.
At small $B_z$, $\ar$ shows a roughly linear increase with approximately the same slope for all $B$ which culminates with a maximum at $B_z \approx 0.5$ T.
In \rfig{fig4}(b) we replot the same data versus $B_z/B$ and observe that the decreasing parts of all curves collapse onto one.
Consistent with \rref{shi:2015a} studying half-integer filling factors in the QH regime, the observed collapse once again confirms that in this parameter range the anisotropy is determined only by the tilt angle.

\begin{figure}[b]
\includegraphics{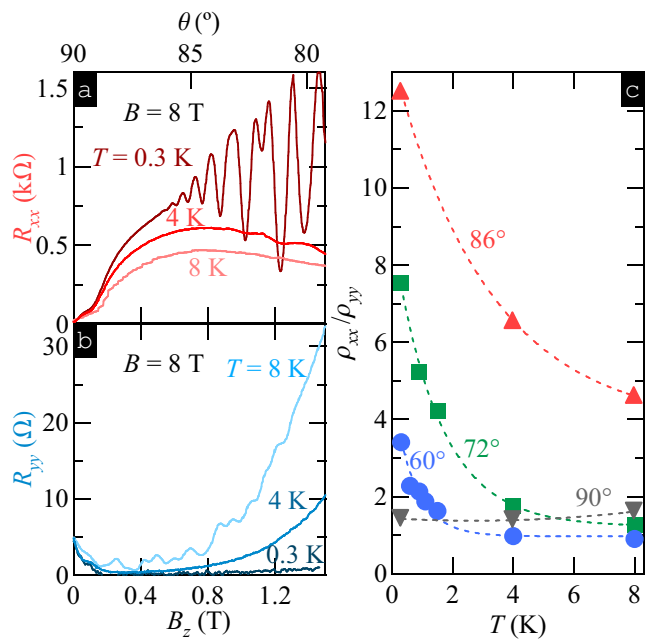}
\caption{(Color online)
(a) $\rxx$ [in k$\Omega$] and (b) $\ryy$ [in $\Omega$] at $B_x \approx B = 8 T$ versus $B_z$, introduced via rotation of the sample, at $T = 0.3, 4$ and 8 K. 
(c) $\rhoxx/\rhoyy$ versus $T$ at $\theta = 60^\circ, 72^\circ, 86^\circ$, and $90^\circ$, as marked. 
Dotted lines are guides for an eye.}
\label{fig5}
\end{figure}

To examine the temperature dependence of the anisotropy in this regime, we present in \rfig{fig5} (a) $\rxx$ and (b) $\ryy$ versus $B_z$ measured at $B_x \approx B = 8$ T and $T =  0.3, 4$, and 8 K.
With increasing temperature, $\rxx$ decreases while $\ryy$ increases, signaling the decrease of the anisotropy over the whole range of $B_z$, except $B_z = 0$.
In \rfig{fig4}(c) we present $\ar$ versus $T$, measured at $\theta = 60^{\circ}, 72^{\circ}, 86^{\circ}$, and $90^{\circ}$, as marked.
We observe that at all tilt angles (except $\theta = 90^\circ$), the anisotropy decays with increasing temperature and that the rate of this decay drops considerably with increasing tilt angle.
Indeed, while at $\theta = 60^\circ$, the anisotropy disappears at $T \approx 2$ K, the resistivity ratio measured at $\theta = 86^\circ$ remains significant, $\rhoxx/\rhoyy > 4$, even at $T = 8$ K. 
At $\theta = 90^{\circ}$, on the other hand, we observe virtually no temperature dependence of the anisotropy.
This finding suggests that the mechanism responsible for the temperature dependence in tilted fields is completely absent in pure $\bip$.

Observation of strong anisotropy in small $B_z$ and at high $Т$ unambiguously rules out QH stripes as a possible origin.
First, the robustness against temperature suggests a much larger energy scale than expected of the charge density wave.
In the Hartree-Fock approach, the latter is similar to the exchange energy \citep{koulakov:1996,fogler:1996}, which is $\sim$ 1 K at $B_z = 0.5$ T.
Indeed, in clean GaAs systems, stripes manifest only at much lower $T$, even in tilted magnetic fields.
Second, QH stripes are expected only when spin-splitting is resolved while in our experiment, at large enough $\bip$, the anisotropy sets in as soon as $B_z$ is added.

The perpendicular magnetic field certainly plays a crucial role in the underlying mechanism of the anisotropy.
Our Ge quantum wells exhibit a modest transport anisotropy both at $B=0$ and in a purely in-plane magnetic field [see \rfig{fig1}(e) and \rref{shi:2015a}].
It is known that an in-plane field could induce anisotropy due to the distortion of the Fermi contour\citep{heisz:1996,dassarma:2000} and surface roughness, via anisotropic, random perpendicular magnetic fields \citep{bykov:2001a,bykov:2001b,goran:2008}.
While the former can be ruled out because it leads to $\rhoxx < \rhoyy$ (when $\bip = B_x$), the latter is consistent with our observations.
We thus conclude that these modest anisotropies likely originate from the surface roughness \citep{hassan:2014}.

One important question is whether the surface roughness can also result in huge anisotropy in our Ge quantum wells in tilted $B$.
First, the $\bip$-induced anisotropy is known to be temperature-independent,\citep{bykov:2001a,goran:2008} whereas the anisotropy in tilted fields has significant temperature dependence [see \rfig{fig5} and \rref{shi:2015a}]. 
Second, because of the anisotropy at $B=0$, the magnitude of the $\bip$-induced anisotropy must depend on the orientation of $\bip$ \citep{goran:2008}, whereas the observed anisotropy in tilted magnetic fields was found to be insensitive to the orientation of $\bip$ \citep{shi:2015a}.
Finally, no strong enhancement of the anisotropy due to additional $B_z$ has been reported in experiments using GaAs samples with much larger surface roughness\citep{bykov:2001a} or predicted theoretically\citep{mirlin:1998b,calvo:1998,evers:1999}.
In fact, \rref{bykov:2001a} reported a \emph{reduction} of the anisotropy upon introduction of $B_z$.
We therefore conclude that the anisotropy in purely parallel magnetic field is unlikely to be related to the strong anisotropy in tilted magnetic fields.

Next we illustrate that, the anisotropy ratio drops sharply in the lower, $N=1$ Landau level and that this drop is most pronounced at low temperatures.
This is in contrast to what we have shown in high Landau levels, that the anisotropy in high-mobility Ge/SiGe quantum wells depends only on the tilt angle, in relatively large perpendicular magnetic fields.

\begin{figure}[t]
\includegraphics{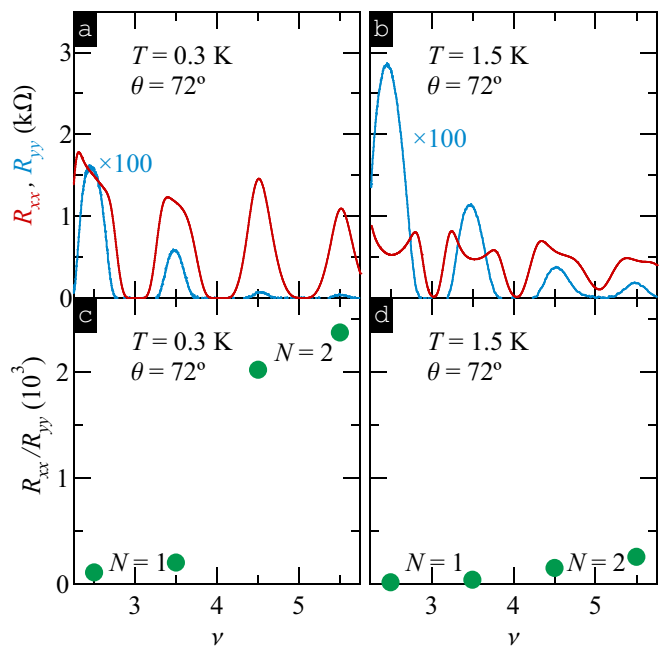}
\caption{(Color online)
$\rxx$ and $\ryy$ [magnified by 100] and their ratio at fixed angle $\theta = 72^\circ$, at (a), (c) $T = 0.3$ K, and (b), (d) $T = 1.5$ K as a function of filling factor $\nu$.
}
\label{fig6}
\end{figure}

In \rfig{fig6}(a) we present $\rxx$ and $\ryy$ (multiplied by 100), measured at $T = 0.3$ K and fixed tilt angle $\theta = 72^{\circ}$, as a function of the filling factor $\nu$.
The data show that while the anisotropy is still present at $\nu = 7/2$ and 5/2, it is considerably weaker than at $\nu = 11/2$ and 9/2.
Indeed, as illustrated in \rfig{fig6}(c), $\rxx/\ryy$ is an order of magnitude smaller at $N = 1$ than at $N = 2$.
\rFig{fig6}(b) shows $\rxx$ and $\ryy$ versus the filling factor measured at the same tilt angle but at $T = 1.5$ K.
We observe that the anisotropy is considerably reduced at this temperature due to both the decrease of $\rxx$ and the increase of $\ryy$.
Interestingly, as shown in \rfig{fig6}(d), the decay of $\rxx/\ryy$ with decreasing $\nu$ becomes much more gradual at $T = 1.5$ K; there is no sharp drop in its value between $N = 2$ to $N =1$, as apposed to $T = 0.3$ K.
In addition, one observes the development of local minima in the $\rxx$ both at $\nu = 5/2$ and at $\nu = 7/2$, and, to a lesser extent, at $\nu = 9/2$ and $\nu = 11/2$.

It is well established that the formation of stripes is less favored in the lower Landau levels due to the smaller number of nodes in the wavefunction.
While the observed decrease of the anisotropy in the $N=1$ Landau level does not have to be related to quantum Hall stripes, the mechanism responsible for a sharp change of the $\rxx/\ryy$ at low temperatures, as shown in \rfig{fig6}(c) between $N$ = 2 and $N$ = 1, can be relevant to the difference in electron wavefunctions.

In summary, we have investigated anisotropic transport in a high-mobility 2D hole gas in a strained Ge quantum well in tilted magnetic fields up to 18 T and at temperatures up to 8 K. 
We have found that the maximum of $\ar$ occurs at the highest available $B_x$ and at $B_z \approx 0.5$ T, where it remains significant even at the highest temperature studied.
The existence of the anisotropy in the regime where no quantum oscillations are seen rules out the formation of stripes as a possible origin.
Even though quantum oscillations are not required, perpendicular magnetic field plays a crucial role both in the magnitude of the anisotropy and its temperature dependence.
We have also shown that our findings are not compatible with other known anisotropies, such as those arising from finite-thickness effects or surface roughness, suggesting a different mechanism of anisotropic transport.

We thank G. Jones, S. Hannas, T. Murphy, J. Park, and D. Smirnov for technical assistance, and A. MacDonald, L. Engel, S. Kivelson,  D. Polyakov, and B. Shklovskii for discussions. 
The work at Minnesota was funded by the US Department of Energy, Office of Basic Energy Sciences, under Grant No. ER 46640 – SC0002567. 
The work at Warwick was supported in part by EPSRC-funded “Spintronic device physics in Si/Ge Heterostructures” EP/J003263/1 and “Platform Grant” EP/J001074/1 projects. 
Q.S. acknowledges the Allen M. Goldman fellowship and the University of Minnesota Doctoral Dissertation Fellowship.
Experiments were performed at the National High Magnetic Field Laboratory, which is supported by NSF Cooperative Agreement No. DMR-0654118, by the State of Florida, and by the DOE.



\end{document}